\documentclass{webofc}

\usepackage[varg]{txfonts}   % Web of Conferences font

\newcommand{\be}{\begin{equation}}
\newcommand{\ee}{\end{equation}}
\newcommand{\Dlt}{\Delta}
\newcommand{\dlt}{\delta}
\newcommand{\prt}{\partial}

\newcommand{\bt}{\beta}
\newcommand{\vp}{\varphi}
\newcommand{\ep}{\varepsilon}
\newcommand{\al}{\alpha}
\newcommand{\ra}{\rightarrow}

\newcommand{\gm}{\gamma}

\newcommand{\dgr}{\dagger}
\newcommand{\lbd}{\lambda}

\newcommand{\rgl}{\rangle}
\newcommand{\lgl}{\langle}

\begin{document}

\title{Critical Temperature in Weakly Interacting Multicomponent Field Theory}

\author{\firstname{V.I.} \lastname{Yukalov}\inst{1}\fnsep\thanks{\email{yukalov@theor.jinr.ru}} \and
        \firstname{E.P.} \lastname{Yukalova}\inst{2}\fnsep\thanks{\email{yukalova@theor.jinr.ru}} 
}

\institute{Bogolubov Laboratory of Theoretical Physics, \\ 
Joint Institute for Nuclear Research, Dubna 141980, Russia 
\and
           Laboratory of Information Technologies, \\ 
Joint Institute for Nuclear Research, Dubna 141980, Russia 
          }

\abstract{
A method is suggested for calculating the critical temperature in
multicomponent field theory with weak interactions. The method is
based on self-similar approximation theory allowing for the
extrapolation of series in powers of asymptotically small coupling
to finite and even infinite couplings. The extrapolation for the
critical temperature employs self-similar factor approximants.
The found results are in perfect agreement with Monte Carlo 
simulations. 
}

\maketitle

\section{Introduction}
\label{sec-1}

Nuclear matter under extreme conditions, when varying baryon density or temperature,
can experience different phase transitions 
\cite{Satz_1,Hagedorn_2,Cleymans_3,Reeves_4,Yukalov_5,Yukalov_6,Yukalov_7}. 
The first thing one needs in studying a phase transition is to find out the 
related transition point, which is not always an easy task. In the present report, 
we explain typical problems arising in calculating transition points by the example 
of an effective multicomponent field theory. The critical temperature in this theory, being 
calculated by means of loop expansions, leads to a series whose terms are divergent. 
So that one has to sum a series of divergent terms, while the resulting temperature 
is, of course, finite. We suggest a method for overcoming this difficulty and 
illustrate it by calculating the critical temperature of an $N$-component field theory 
for different $N$. Comparing our approach with Monte Carlo simulations, when these 
are available, we find that our method provides the results that are in beautiful 
agreement with the Monte Carlo data.

We emphasize that our aim in this report is not to consider a particular phase 
transition that could occur in nuclear matter under varying parameters, but to develop 
a general strategy that would be valid for calculating critical temperatures in any of 
phase transitions of second-order type. As examples of such critical phase transitions 
in nuclear matter, occurring under varying barion density or temperature, it is possible 
to mention the transition at the critical point between hadron gas and hadron liquid, the 
transitions between hadron liquid and nuclear superfluid, between quark-gluon plasma 
and colour superconductor, Bose condensation of multiquark clusters, possible transitions
between gluon and gluball phases, and some scenarios of deconfinement transition    
\cite{Satz_1,Hagedorn_2,Cleymans_3,Reeves_4,Yukalov_5,Yukalov_6,Yukalov_7}.
In the present report, we treat a model possessing those problems that are typical in
calculating critical temperatures in practically any continuous phase transition.   

In what follows, we employ the system of units, where the Planck and Boltzmann
constants are set to unity. 

We start with a nonrelativistic $3 + 1$ - dimensional field theory with the action
\be
\label{A1}
 S[\psi] = \int L[\psi]\; dt \; , \qquad
L[\psi] = \int \psi^\dgr i \; \frac{\prt}{\prt t} \; \psi \; dx \; - \; 
H[\psi] \;  ,
\ee
whose Lagrangian contains the energy part
\be
\label{A2}
 H[\psi] = \int \left [ \psi^\dgr \left ( - \frac{\nabla^2}{2m} - \mu \right ) \psi
+ \frac{\lbd}{2} \; | \psi|^4 \right ] \; dx \; ,
\ee
with the coupling parameter
\be
\label{A3}
\lbd = 4\pi \; \frac{a_s}{m}
\ee
expressed through the scattering length $a_s$.  The term $|\psi|^4$ is understood 
as $(\psi^\dgr \psi)^2$.  In the three-dimensional space, the variable 
$x = \{x_{\alpha}: \alpha = 1,2,3\}$. 

Being interested in static quantities at finite temperature, we pass to the imaginary 
time formalism by accomplishing the Wick rotation \cite{Kleinert_40}. Then the field 
$\psi$ is periodic with respect to the imaginary time, $\psi(x,0) = \psi(x,\beta)$, 
where $\beta \equiv 1/T$ is inverse temperature. Due to this periodicity, the field 
$\psi$ may be decomposed into imaginary-time frequency modes with Matsubara 
frequencies $2 \pi T j$, where the integer $j = 0, \pm 1, \pm 2, \ldots$. Near the 
transition point, the main contribution to the partition function is played by the zero 
Matsubara mode, while the nonzero Matsubara frequencies can be integrated out. 
To take into account $N$ internal degrees of freedom, the term $|\psi|^2$ is interpreted 
as $\sum_{n=1}^N \varphi^2_n$, with the field $\varphi_n$ for each of the components.  
In this way, we come to the partition function being the trace of  the Boltzmann weight 
$\exp S[\varphi]$, with the effective action 
\be
\label{A4}
S[\vp] =  \int \left [   \frac{1}{2} \left ( \frac{\prt\vp}{\prt x} \right )^2
+ \frac{m_{eff}^2}{2}\; \vp^2 + \frac{\lbd_{eff}}{4!}\; \vp^4 \right ] \; dx \; ,
\ee 
corresponding to an $N$-component field theory, with the effective parameters
\be
\label{A5}
 m_{eff}^2 = - 2m T \mu \; , \qquad  \lbd_{eff} = 48 \pi m T a_s \; .
\ee
The details of the above procedure are well known and were described in many
publications, e.g., in Refs. \cite{Kleinert_40,Baym_36,Baym_37,Holzmann_38,Zinn_39}. 

Thus, we need to study phase transitions in the $N$-component filed theory with 
action (\ref{A4}). This action represents a wide class of physical systems, depending
on the number of components. For instance, $N = 0$ corresponds to dilute polymer
solutions, $N = 1$, to the Ising-type models, $N = 2$, to the XY model and superfluids,
and $N = 3$ corresponds to the Heisenberg model.

\section{Phase transition temperature}
\label{sec-2}
 
Action (\ref{A4}) describes the $O(N)$-symmetric multicomponent filed theory
with the $N$-component field $\varphi = \{\varphi_n(x): n = 1,2,\ldots,N\}$. The 
action is invariant under the inversion symmetry
\be
\label{2}
\vp_n \ra - \vp_n \qquad ( n = 1,2, \ldots, N) \;   .
\ee
This implies that one of the solutions for the order parameter $\eta\equiv\lgl\vp\rgl$, 
which is the average of the field, is zero,  $\langle\varphi\rangle = 0$. 

But below the critical temperature $T_c$, the symmetry can become spontaneously 
broken, when a nonzero order parameter $\langle \varphi \rangle$ provides a lower 
free energy  $F = - T \ln {\rm Tr}\exp(-S[\vp])$, so that
\be
\label{3}
 F(\eta\neq 0) < F (\eta\equiv 0) \;  .
\ee
One says that at the critical temperature $T_c$ there occurs a phase transition, where
the order parameter changes from zero to a nonzero value:
\be
\label{4}
\eta \equiv 0 \qquad ( T > T_c ) \; ; \qquad  \eta \neq 0 \qquad ( T < T_c ) \;  .
\ee
In the considered case, the order parameter at $T_c$ is continuous, which signifies
the second order phase transition.

The transition temperature depends on the interaction strength that can be characterized
by the gas parameter
\be
\label{5}
 \gm \equiv \rho^{1/3} a_s \;  ,
\ee
in which $\rho$ is a particle density and the scattering length is assumed to be positive, 
$a_s > 0$. For the ideal gas, with no interactions, the critical temperature is
\be 
\label{6}
 T_0 \equiv T_c(0) = 
\frac{2\pi}{m} \left [ \frac{\rho}{\zeta(3/2)} \right ]^{2/3} \;  ,
\ee
which is evident from Eqs. (\ref{A1}) and  (\ref{A2}) under $\lambda = 0$. The 
problem is to find the dependence of the critical temperature $T_c(\gamma)$ 
on the gas parameter at small values of the latter. 

One defines the relative critical temperature shift
\be
\label{7}
\frac{\Dlt T_c}{T_0} \equiv \frac{T_c - T_0}{T_0} \;   ,
\ee
which, at weak interactions, is linear in the gas parameter:
\be
\label{8}
 \frac{\Dlt T_c}{T_0} \simeq c_1 \gm \qquad ( \gm\ra 0) \;  ,
\ee
as it has been proved in Refs. \cite{Baym_36,Baym_37,Holzmann_38,Zinn_39}. 
The difficulty in finding the coefficient $c_1$ is in the impossibility of employing
perturbation theory with respect to $\gamma$, which, as explained below, looses 
its sense at the point $T_c$. Various ways of solving this problem have been 
reviewed in \cite{Andersen_8,Yukalov_9}.
 
One is able to calculate the shift coefficient $c_1$ involving the loop expansion
\cite{Kleinert_10}, which yields \cite{Kastening,Kastening_12,Kastening_13} an 
expansion in powers of the variable
\be
\label{9}
 x = ( N+2) \; \frac{\lbd_{eff}}{\sqrt{\mu_{eff}} } \;  , 
\ee
in which $N$ is the number of the components, $\lambda_{eff}$ is an effective 
coupling, defined in (\ref{A5}), and $\mu_{eff}$ is
\be
\label{1}
 \mu_{eff} = m^2_{eff} - \Sigma(0,0)   \; ,
\ee
with  $\Sigma(0,0)$ being the self-energy at zero limit of energy and momentum. 

In seven loops, one comes to the asymptotic expansion
\be
\label{10}
 c_1(x) \simeq \sum_{n=1}^5 a_n x^n \qquad ( x\ra  0) \;  .
\ee

The problem arises, when one considers the above expansion at the critical temperature
$T_c$, where $\mu_{eff}$ becomes zero, because of which the expansion variable
tends to infinity, $x \ra \infty$. Since this expansion has been derived for asymptotically 
small $x$, the tendency of the latter to infinity makes the expansion senseless. To make 
sense of such a difficulty, it is necessary to invent  a way of finding an effective sum of 
the asymptotic series, derived for $x \ra 0$, which would extrapolate the sum to finite 
values of $x$ and including $x \ra \infty$.      

It is worth mentioning that the extrapolation of asymptotic series for a small variable
to arbitrary values of the latter can be accomplished in the frame of optimized 
perturbation theory \cite{Yukalov_14,Yukalov_15}. Different variants of this theory have 
been employed for finding the shift coefficient $c_1$ by introducing control functions with
a variable change \cite{Kastening_12,Kastening_13} or incorporating them into initial 
approximations \cite{Souza_17,Souza_18,Kneur_20,Kneur_21}. However, the 
results of such calculations strongly depend on how the control functions are
introduced, and also these calculations are rather cumbersome requiring the use of 
optimization conditions defining control functions at each approximation order.  

Below we suggest a general uniquely defined method for extrapolating asymptotic series
of a small variable to arbitrary values of the variable, including its infinite limit.

\section{Self-similar approximation theory}
\label{sec-3}

A general method of extrapolation is based on self-similar approximation theory
\cite{Yukalov_22,Yukalov_23,Yukalov_24,Yukalov_25}. Although the final prescription 
that will be formulated below is rather simple, we briefly delineate the main steps 
of the theory to give the reader the impression of how the employed approximants 
have been obtained.   

Self-similar approximation theory is based on the following important points. 
First, it is necessary to transform an initially given divergent sequence 
$\{f_k(x)\}$ of the series of order $k$ for the sought function into a convergent 
sequence $\{F_k(x,u_k(x))\}$ by incorporating control functions $u_k(x)$. The latter 
can be introduced either in an initial approximation $f_0(x,u)$, or through a 
variable change $x=z(x,u)$, or by a sequence transformation $T(u)f_k(x)=F_k(x,u)$. 
A sequence is convergent if and only if, for a given positive $\varepsilon$, there 
exists $k_\varepsilon$ such that for all $k>k_\varepsilon$ the Cauchy criterion 
of convergence is valid:
\be
\label{11}
 | F_{k+p}(x,u_{k+p}) - F_k(x,u_k) |  < \ep \;  .
\ee
Control functions can be defined by minimizing the Cauchy cost functional
\be
\label{12}      
 C[u] = \frac{1}{2} \sum_k | F_{k+p}(x,u_{k+p}) - F_k(x,u_k) |^2 \;  .
\ee

At the next step, we consider the passage from an approximation $F_k$ to $F_{k+1}$
and so on as the motion in discrete time $k$. To formulate this in terminology of 
dynamical theory, we impose the reonomic constraint
\be
\label{13}
F_0(x,u_k(x) ) = f \; , \qquad x = x_k(f)
\ee
and introduce the endomorphism
\be
\label{14}
 y_k(f) \equiv F_k(x_k(f),u_k(x_k(f))) \;  .
\ee
In order for the Cauchy cost functional (\ref{12}) to have the absolute minimum 
zero, it is necessary that endomorphism (\ref{14}) would satisfy the relation
\be
\label{15}
y_{k+p}(f) = y_k(y_p(f) )
\ee
that is termed the property of {\it functional self-similarity}. Relation (\ref{15}) 
defines a dynamical system in discrete time, that is, a cascade, whose trajectory is 
bijective to the approximation sequence $\{F_k\}$, because of which this dynamical 
system is called the {\it approximation cascade}.  

It is more convenient to deal with a dynamical system in continuous time, i.e., 
a flow. For this purpose, it is possible to embed the approximation cascade into 
an {\it approximation flow}, which implies that the flow trajectory passes through 
all points of the cascade trajectory,
\be
\label{16}
\{ y_k(f): \; k \in \mathbb{Z}_+ \} \subset \{ y_t(f): \; t \in \mathbb{R}_+ \} \;  .
\ee
For the approximation flow, the self-similarity relation (\ref{15}) can be rewritten 
as the Lie equation 
\be
\label{17}
 \frac{\prt}{\prt t} \; y_t(f) = v(y_t(f)) \; , \qquad   
v(f) \equiv \left [ \frac{\prt}{\prt t} \; y_t(f) \right ]_{t=0} \;  .
\ee
The Lie equation of motion, being integrated, yields the evolution integral
\be
\label{18}
 \int_{y_k}^{y_k^*} \frac{dy}{v_k(y)} = t_k \;  ,
\ee
in which  $v_k$ is the cascade velocity that is the Euler discretization of the flow 
velocity, $y_k^*$ is a quasi-fixed point, and $t_k$ is an effective time required for
reaching $y_k^*$ from $y_k$. The quasi-fixed point $y_k^*$ represents the effective
limit of the flow trajectory, hence, being bijective to the approximation sequence 
through notation (\ref{14}), it represents the self-similar approximation $f_k^*(x)$
for the sought function $f(x)$. 

The stability of the approximation procedure can be controlled by studying the local
map multipliers
\be
\label{19}
 \mu_k(f) \equiv \frac{\dlt y_k(f)}{\dlt y_{k-1}(f)} \;  .
\ee
The motion is locally stable provided that the absolute values of the map multipliers
are not larger than one. Approaching a fixed point, one has 
$|\mu_k(f) | \ra 1$ as $y_k \ra y^*$. More details on the self-similar approximation 
theory can be found in the papers
\cite{Yukalov_22,Yukalov_23,Yukalov_24,Yukalov_25} and summarized in the 
review papers \cite{Yukalov_27,Yukalov_28,Yukalov_29}. 

Let us concretize the approach by considering a function for which one can find only
an asymptotic series at a small variable,
\be
\label{20}
 f(x) \simeq f_k(x) \qquad ( x \ra 0 ) \;  ,
\ee
with the $k$-th order expansion
\be
\label{21}
 f_k(x) = f_0(x) \left ( 1 + \sum_{n=1}^k a_n x^n \right ) \;  ,
\ee
in which $f_0(x)$ is a known term.  Applying a fractal transform  \cite{Yukalov_28} 
to series (\ref{21}) and following the procedure delineated above, we come to the 
self-similar factor approximants
\be
\label{22}
 f_k^*(x) = f_0(x) \prod_{i=1}^{N_k} ( 1 + A_i x)^{n_i} \;  ,
\ee
where $N_k = k/2$ for even $k = 2,4,\ldots$ and $N_k = (k+1)/2$ for odd $k = 3,5,\ldots$. 
The parameters $A_i$ and $n_i$ are defined by the accuracy-through-order procedure
$ f_k^*(x) \simeq f_k(x)$ as $x \ra 0$, with the scaling $A_1 = 1$ for odd approximants 
\cite{Yukalov_30,Gluzman_31,Yukalov_32,Yukalov_33}. 

We aim at studying the limit of large $x$. Let the known term $f_0(x)$ behave as
\be
\label{23}
 f_0(x) \simeq A x^\al \qquad ( x\ra\infty ) \;  .
\ee
If the large-variable limit of the sought function is 
\be
\label{24}
 f(x) \simeq B x^\bt \qquad ( x\ra\infty ) \;  ,
\ee
then the powers of approximant (\ref{22}) have to obey the constraint
\be
\label{25}
 \al + \sum_{i=1}^{N_k} n_i = \bt \qquad ( \bt \neq 0) \;  ,
\ee
which, for the case of a finite large-variable limit, reduces to
\be
\label{26}
\al + \sum_{i=1}^{N_k} n_i = 0 \qquad ( \bt = 0) \;    .
\ee
 
In this way, we get a sequence of self-similar factor approximants (\ref{22}). If 
a number of terms in sum (\ref{21}) are available, then all we need is to observe 
the numerical convergence of the sequence $\{f_k^*\}$. When there are a few of such 
terms, the convergence  of self-similar approximants can be essentially accelerated 
by constructing a quadratic spline
\be   
\label{27}
 q(x,t) = b_0(x) + b_1(x) t + b_2(x) t^2 \;  ,
\ee
whose coefficient functions are defined through the conditions
\be
\label{28}
 q(x,0) = f_{k-2}^*(x) \; , \qquad  q(x,1) = f_{k-1}^*(x) \; , \qquad 
q(x,2) = f_{k}^*(x) \; .
\ee
The self-similar approximant for this spline reads as
\be
\label{29}
 q^*(x,t) = b_0(x) [ 1 + A(x) t]^{n(x)} \;  .
\ee
The latter provides an extrapolation of the sequence $f_{k-2}^*$,  $f_{k-1}^*$, and 
$f_k^*$ to larger orders by setting $t \geq 2$. The minimal extrapolation involves 
$t = 2$ and $t = 3$ yielding the doubly renormalized approximant
\be
\label{30}
f_k^{**}(x) = \frac{1}{2} \; \left [ q^*(x,2) + q^*(x,3) \right ] \; ,
\ee
whose error value is given by the difference of $q^*(x,2)$ and  $q^*(x,3)$. 
    
\section{Calculation of temperature shift}
\label{sec-4}

We accomplish this procedure for the critical temperature shift (\ref{8}) using 
the fifth-order series (\ref{10}). The coefficients $a_n$, calculated using the 
data of the seven-loop expansion by Kastening \cite{Kastening_12,Kastening_13}, 
for different numbers of the components $N$, are presented in Table 1. 

Note that in expansion (\ref{10}) the variable $x$ is assumed to be small, such
that $x \ra 0$, independently of the values of the parameters entering notation (\ref{9}).
But at the end, we have to find the limit $x \ra \infty$, because of which the final 
answer for the coefficient $c_1$ is $c_1 = \lim_{x\ra\infty} c_1^{**}(x)$. 

Following the scheme of the previous section, we find the $k$-order factor 
approximants $f_k(x)^*$ for the quantity $c_1(x)$ in third order,
$$
 f_3^*(x) = a_1 x (1 + x)^{n_1}  (1 + A_2 x)^{n_2} \; ,
$$
with $n_1 + n_2 = -1$,  in fourth order,
$$
 f_4^*(x) = a_1 x (1 + A_1 x)^{n_1}  (1 + A_2 x)^{n_2} \;  ,
$$
where  $n_1 + n_2 = -1$, and in fifth order,
$$
f_5^*(x) = a_1 x (1 + x)^{n_1} (1 + A_2 x)^{n_2} (1 + A_3 x)^{n_3} \;   ,
$$
with  $n_1 + n_2 + n_3 = -1$. 

Since finally we need the limits $x \ra \infty$, we find from the above approximants
the limits 
$$
 f_k^* \equiv \lim_{x\ra\infty} f_k^*(x)  
$$
in the forms
$$
f_3^* = a_1 A_2^{n_2} \; , \qquad  f_4^* = a_1 A_1^{n_1} A_2^{n_2} \; , \qquad 
 f_5^* = a_1 A_2^{n_2} A_3^{n_3} \;  ,
$$
whose values are given in Table 2. After taking the limit $x \ra \infty$, the variable $x$
does not enter the expressions below.   

Accelerating the convergence by means of splines, as is explained above, we construct 
the quadratic polynomials
$$
 q(t) = b_0 + b_1 t + b_2 t^2 \;  ,
$$
with the coefficients defined by the equations
$$
q(0) = f_3^* \; , \qquad  q(1) = f_4^* \; , \qquad q(2) = f_5^* \; , 
$$
which are listed in Table 3. The factor approximant for the spline $q(t)$ reads as
$$
 q^*(t) = b_0 (1 + At)^n \;  ,
$$
with the values $A$ and $n$ from Table 3. The values
$$
q_2^* \equiv q^*(2) \; , \qquad q_3^* \equiv q^*(3)
$$
are presented in Table 2. 

Finally, the sought coefficient $c_1$, defined by formula (\ref{30}), under $x \ra \infty$ 
is given in Table 4. We also compare our calculations for $c_1$ with the available 
Monte Carlo simulations \cite{Kashurnikov_34,Arnold_36,Arnold_37,Sun_38},
as well as with the results obtained by optimized perturbation theory (OPT) 
\cite{Kastening_12,Kastening_13}. As it is seen, the values of $c_1$, derived in 
our approach, are in perfect agreement with the Monte Carlo simulations. The 
advantage of the self-similar factor approximants, we employed, is that they 
are uniquely defined and do not involve the use of control functions at the 
final step. While the results of optimized perturbation theory strongly depend 
on the choice of such control functions.

\section{Discussion}
\label{sec-5}

We have suggested a method of extrapolating asymptotic series for a small 
variable $x \ra 0$ to any finite values of $x$, including the infinite limit 
$x \ra \infty$. The method is based on self-similar approximation theory. 
We have applied the method to calculating the critical temperature of weakly 
interacting $O(N)$ symmetric field theory. We have considered the cases with 
the number of the components $N = 0,1,2,3,4$, for which the seven-loop 
expansions are known. Note that the case of $N = \infty$ allows for an 
exact solution \cite{Baym_37,Arnold_40} yielding
$$
c_1(N=\infty) = \frac{8\pi}{3[\zeta(3/2)]^{4/3}} = 2.328473 \; .
$$

As is evident from (\ref{7}) and (\ref{8}), knowing $c_1$ immediately gives 
the critical temperature
$$
T_c  \simeq T_0 (1 +  c_1 \gamma). 
$$ 

Our results are in very good agreement with Monte Carlo simulations. The 
advantage of the used self-similar factor approximants is in their simplicity 
and unique definition, involving no control functions in the final calculations.

\begin{acknowledgement} 
We are grateful to B. Kastening for discussions. Financial support from RFBR 
(grant $\# 14-02-00723$) is acknowledged.
\end{acknowledgement}

\begin{table}
\centering
\caption{The coefficients $a_n$ of the asymptotic expansion for $c_1(x)$.}
\label{tab-1}    
\begin{tabular}{|c|c|c|c|c|c|} \hline
$N$   &         0    &       1      &    2          &    3    &    4     \\ \hline
$a_1$ &    0.111643  &    0.111643  &    0.111643   &    0.111643   &    0.111643 \\ \hline
$a_2$ & $-$0.0264412 & $-$0.0198309 & $-$0.0165258  & $-$0.0145427  & $-$0.0132206 \\ \hline
$a_3$ &    0.0086215 &    0.00480687&    0.00330574 &    0.00253504 &    0.0020754 \\ \hline
$a_4$ & $-$0.0034786 & $-$0.00143209& $-$0.000807353& $-$0.000536123& $-$0.000392939 \\ \hline
$a_5$ &    0.00164029&    0.00049561&    0.000227835&    0.000130398&    0.0000852025 \\ \hline
\end{tabular}
\end{table}

\begin{table}
\centering
\caption{Self-similar factor approximants $f_k^*(\infty),\; k=3,4,5$, and $q_m^*,\; m=2,3$.}
\label{tab-2}    
\begin{tabular}{|c|c|c|c|c|c|} \hline
$N$     &       0    &    1      & 2        &    3    & 4       \\ \hline
$f^*_3$ &  0.548162  & 0.700831  & 0.823464 & 0.92360 & 1.00681   \\  \hline
$f^*_4$ &  0.646479  & 0.858742  & 1.02476  & 1.15706 & 1.26426  \\ \hline
$f^*_5$ &  0.682986  & 0.888388  & 1.05111  & 1.18101 & 1.28629  \\  \hline
$q^*_2$ &  0.7347    & 1.008     & 1.219    & 1.384   & 1.516   \\  \hline
$q^*_3$ &  0.7999    & 1.112     & 1.352    & 1.540   & 1.688   \\  \hline
\end{tabular}
\end{table}

\begin{table}
\centering
\caption{Coefficients of polynomials $q(t)=b_0+b_1t+b_2t^2$, and
self-similar approximants $q^*(t) = b_0(1+At)^n$.}
\label{tab-3}    
\begin{tabular}{|c|c|c|c|c|c|} \hline
$N$   &     0       &       1      &    2         &    3        &    4     \\ \hline
$b_0$ & 0.548162    & 0.700831     & 0.823464     & 0.9236      & 1.00681    \\ \hline
$b_1$ & 0.129222    & 0.222044     & 0.288772     & 0.338205    & 0.375156 \\  \hline
$b_2$ & $-$0.030905 & $-$0.0641325 & $-$0.087474  & $-$0.104749 & $-$0.11771 \\ \hline
$A$   & 0.714061    & 0.894486     & 0.956514     & 0.985622    & 1.00014   \\  \hline
$n$   & 0.330135    & 0.354202     & 0.366623     & 0.371523    & 0.372564  \\  \hline
\end{tabular}
\end{table}

\begin{table}
\centering
\caption{Results for the coefficient $c_1$ of the critical temperature shift 
obtained using self-similar factor approximants, compared to Monte Carlo 
simulations and optimized perturbation theory.}
\label{tab-4}
\begin{tabular}{|c|c|c|c|} \hline
$N$ &     $c_1$      &       Monte Carlo                    &    OPT         \\ \hline
0   & 0.77$\pm$ 0.03 &                                      & 0.81$\pm$ 0.09 \\ \hline
1   & 1.06$\pm$ 0.05 & 1.09$\pm$ 0.09  \cite{Sun_38}        & 1.07$\pm$ 0.10 \\ \hline
2   & 1.29$\pm$ 0.07 & 1.29$\pm$ 0.05 \cite{Kashurnikov_34} & 1.27$\pm$ 0.11  \\
    &                & 1.32$\pm$ 0.02 \cite{Arnold_36,Arnold_37} &            \\ \hline
3   & 1.46$\pm$ 0.08 &                     & 1.43$\pm$ 0.11  \\ \hline
4   & 1.60$\pm$ 0.09 & 1.60$\pm$ 0.10 \cite{Sun_38} & 1.54$\pm$ 0.11  \\ \hline
\end{tabular}
\end{table}

\end{document}